# Ultra high stiffness and thermal conductivity of graphene like $C_3N$

Bohayra Mortazavi

*Institute of Structural Mechanics, Bauhaus-Universität Weimar,*

*Marienstr. 15, D-99423 Weimar, Germany*

**Abstract**

Recently, single crystalline carbon nitride 2D material with a $C_3N$ stoichiometry has been synthesized. In this investigation, we explored the mechanical response and thermal transport along pristine, free-standing and single-layer $C_3N$. To this aim, we conducted extensive first-principles density functional theory (DFT) calculations as well as molecular dynamics (MD) simulations. DFT results reveal that $C_3N$ nanofilms can yield remarkably high elastic modulus of 341 GPa.nm and tensile strength of 35 GPa.nm, very close to those of defect-free graphene. Classical MD simulations performed at a low temperature, predict accurately the elastic modulus of 2D $C_3N$ with less than 3% difference with the first-principles estimation. The deformation process of $C_3N$ nanosheets was studied both by the DFT and MD simulations. Ab initio molecular dynamics simulations show that single-layer $C_3N$ can withstand high temperatures like 4000 K. Notably, the phononic thermal conductivity of free-standing $C_3N$ was predicted to be as high as 815±20 W/mK. Our atomistic modelling results reveal ultra high stiffness and thermal conductivity of $C_3N$ nanomembranes and therefore propose them as promising candidates for new application such as the thermal management in nanoelectronics or simultaneously reinforcing the thermal and mechanical properties of polymeric materials.

Corresponding author (BohayraMortazavi):  bohayra.mortazavi@gmail.com

Tel: +49 176 68195567;

Fax: +49 364 358 4511

## 1. Introduction

Since 2004 that the first mechanical exfoliation of graphene [1–4] from graphite was achieved successfully, two-dimensional (2D) materials with only few atomic layers thickness have emerged as a new class of materials. In particular, graphene has been so-far acting a unique role as the most prominent member of 2D materials, owing to its superior thermal conductivity [5,6], mechanical [7] and electronic [4] properties. After the great success of graphene, single-layer forms of the other members of the group of IV elements so called silicene [8,9], germanene [10] and stanene [11], have played a significant role in the field of 2D materials. Nevertheless, likely to graphene in its free-standing form, silicene, germanene and stanene present zero-band-gap semiconducting electronic character, which limit their suitability for particular applications. During the last decade, this limitation of the graphene however has been acting as one of the strongest motivations for the experimental researches, in order to synthesize new 2D materials with inherent semiconducting electronic properties. Amazingly, experimental advances could establish practical routes toward the fabrication of a wide range of high-quality 2D crystals such as hexagonal boron-nitride (h-BN) [12,13], phosphorene [14,15], graphitic carbon nitride [16,17] and transition metal dichalcogenides like $MoS_2$ and $WS_2$ [18,19].

Covalent networks of the carbon and nitrogen atoms, have attracted remarkable attention as a new class of 2D materials with semiconducting electronic properties, well suited for a wide range of applications. Graphitic carbon nitride, $g-C_3N_4$, structures have been synthesized for a long-period by polymerization of cyanamide, dicyandiamide or melamine [16]. Graphitic carbon nitride structures have shown great potential applications for energy conversion and storage and environmental applications such as direct methanol fuel cells, catalysis, photocatalysis and $CO_2$ capture [16,20–25]. Nevertheless, large area and high quality triazine-based covalently-linked, $sp^2$-hybridized carbon and nitrogen atoms, with only few atomic layer thickness and semiconducting electronic properties, have been fabricated recently using an ionothermal, interfacial reaction [17]. Both experimental measurements and theoretical calculations [17], confirmed that triazine-based graphitic carbon nitride 2D materials can present a direct band-gap between 1.6 eV and 2.0 eV, which accordingly highlights their promising applications in nanoelectronics, as field-effect transistors or light-emitting diodes. In line with continuous efforts for the synthesis of novel 2D materials consisting of carbon and nitrogen covalent networks, in 2015 nitrogenated holey graphene (NHG) with ordered distributed holes and nitrogen atoms and a $C_2N$ stoichiometry were successfully synthesized via a simple wet-chemical reaction [26]. As expected and likely to triazine-based graphitic carbon nitride, nitrogenated holey graphene also yields band-gaps of approximately 1.70 eV and 1.96 eV [26]. The interest toward the fabrication of 2D materials made only from carbon and nitrogen atoms seems to be highly attractive, and such that an exciting experimental advance has just taken place with respect to the synthesis of 2D polyaniline crystals with $C_3N$ stoichiometry [27]. 2D polyaniline lattice is analogous to that of the defect-free graphene, which contains uniformly distributed nitrogen atoms with an ordered pattern.

Successful experimental synthesis of 2D polyaniline $C_3N$ nanomembranes consequently raise the importance of the evaluation of their intrinsic properties. Such that, comprehensive understanding of the thermal, mechanical, optical and electronic properties of 2D polyaniline not only plays a crucial role in their usage in nanodevices but also may propose them as suitable candidates for new applications. Because of the difficulties and complexities of the experimental characterizations for 2D materials with only few-atomic layer thickness, theoretical methods can be considered as promising alternatives to investigate their properties [28–34]. In this investigation, we therefore conducted extensive atomistic modelling to evaluate the mechanical properties and thermal conductivity of single-layer and free-standing 2D polyaniline $C_3N$.

## 2. Methods

The atomic structure of 2D polyaniline $C_3N$ structure is illustrated in Fig. 1. $C_3N$ can be considered as a nitrogen doped graphene in which the nitrogen atoms substitute the native carbon atoms in the pristine graphene, in an ordered pattern. Likely to graphene, $C_3N$ also presents two major orientations, so called armchair and zigzag directions as depicted in Fig. 1. In this study, we explored the mechanical and thermal properties of $C_3N$ along the both armchair and zigzag directions. The mechanical responses of free-standing $C_3N$ were analyzed by the first-principles density functional theory (DFT) calculations and classical molecular dynamics simulations as well. The thermal conductivity of single-layer $C_3N$ was however only estimated using the non-equilibrium molecular dynamics simulations.

The DFT calculations in this study were performed using the Vienna ab initio simulation package (VASP) [35–37]. The plane wave basis set with an energy cutoff of 500 eV and the gradient approximation exchange-correlation functional potential, formulated by Perdew-Burke-Ernzerhof [38], was employed. A rectangular super-cell consisting of 32 atoms was constructed with dimensions of 8.42 Å ×9.72 Å. We applied the periodic boundary conditions in all three Cartesian directions to avoid the effect of free-atoms on edges. We additionally considered a vacuum layer of 20 Å to avoid image-image interactions along the single-layer $C_3N$ normal direction. We used 15×15×1 Monkhorst-Pack [39] k-point mesh size for the initial minimization of the structures. In this case, the simulation box size was allowed to change to ensure no residual stresses in the structure. We used $10^{-5}$ eV criteria for the energy convergence and 0.005 eV/Å threshold for the forces. After obtaining the minimized structure, we applied uniaxial tension loading to evaluate the mechanical properties of single-layer $C_3N$. To do so, we increased the periodic simulation box size along the loading direction in a step by step procedure, every step with a small strain of 0.002.

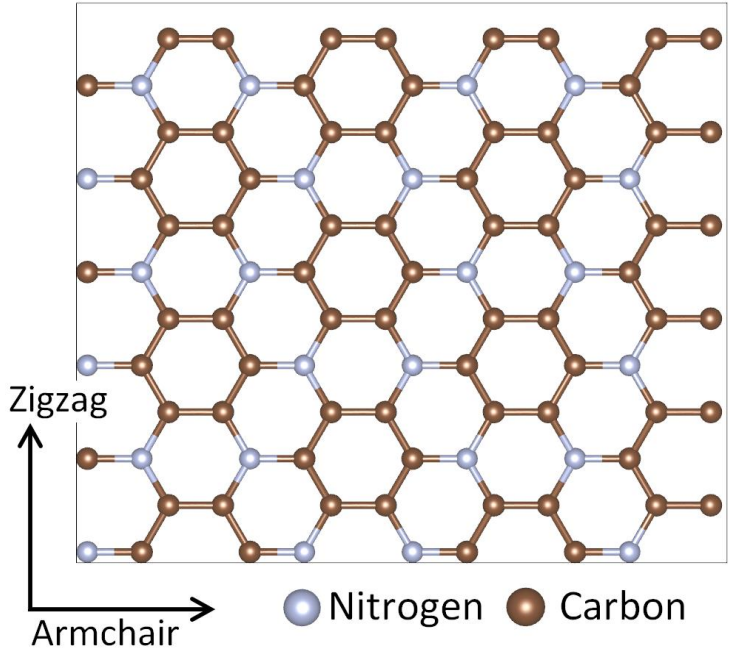

Fig.1- Atomic structure of 2D polyaniline $C_3N$. To evaluate the mechanical properties using the DFT method, we constructed a rectangular super-cell with 32 atoms with dimensions of 8.42 Å×9.72 Å.

Due to the fact that we studied a single-layer structure, the atoms are in contact with vacuum along the sheet's normal direction and such that the stress along the sheet thickness is negligible. However, as a result of applied strain along the loading direction, the in-plane stress perpendicular of the loading direction may not be zero. Therefore, to ensure accurate uniaxial stress condition, the simulation box size in the perpendicular direction of the loading was altered, with a goal to reach the negligible stress in this direction. We note that after applying the changes in the simulation box size, the atomic positions were consistently rescaled and therefore no sudden void formation or bond stretching in the lattice were occurred. We than used the conjugate gradient method for the geometry optimizations, with termination criteria of $10^{-5}$ eV and 0.01 eV/Å for the energy and the forces, respectively, by employing a 5×5×1 Monkhorst-Pack k-point mesh size. To evaluate the electron localization function and electronic density of states (DOS) using the PBE method, we performed a single-point calculation with a denser k-point mesh of 15×15×1. We also calculated the electronic DOS using HSE06 [40] hybrid functional with 8×8×1 k-point mesh. To investigate the thermal stability of the single-layer $C_3N$, we performed ab initio molecular dynamics (AIMD) simulations, using the Langevin thermostat with a time step of 1fs and 2×2×1 k-point mesh. Based on our DFT results, the carbon-carbon and carbon-nitrogen bond lengths were found to be very close, 1.404 Å and 1.403 Å, respectively. To plot the VASP results, we used VESTA [41] package.

Molecular dynamics (MD) simulations in the work were carried out using the LAMMPS (*Large-scale Atomic/Molecular Massively Parallel Simulator*) [42] package. Tersoff potential [43,44] was used for introducing the atomic interactions. We used the optimized Tersoff potential proposed by the Lindsay and Broido [45] for introducing the carbon atoms interactions. Tersoff potential parameters for the carbon-nitrogen interactions were adopted from the work by Kinarci *et al.* [46]. The accuracy of the predictions derived from the MD simulations, strongly correlates to the appropriate selection of the forcefield to define the atomic interactions. It is worthy to note that the optimized Tersoff potential [45], to the best of our knowledge is currently the most accurate choice for the MD simulation of the thermal transport along $sp^2$ carbon structures, mainly because it reproduces the phonon dispersion curves of graphite in a close agreement with experimental measurements. In addition, carbon-nitrogen Tersoff potential parameters set, proposed by the Kinarci *et al.* [46], was particularly developed to investigate the thermal transport. Based on our MD geometry optimization, the carbon-carbon and carbon-nitrogen bond lengths were found to be 1.44 Å and 1.43 Å, respectively. These bond lengths are in less than 3% difference with our DFT predictions. The mentioned augments reveal that our choice of potential functions for the MD modelling of thermal transport along 2D $C_3N$ is theoretically convincing.

In our MD modelling and in agreement with our DFT calculations, we performed uniaxial tensile simulations to investigate the mechanical properties. In this case, we applied the periodic boundary conditions along the planar directions and a small simulation time step of 0.25 fs was adopted. Moreover, the simulation box for the uniaxial simulations included around 14,000 individual atoms. First, the structure was relaxed to zero stresses along the planar directions and at the desired temperature using the Nosé-Hoover barostat and thermostat (NPT) method. Next, the periodic simulation box size along the loading direction was increased by a constant engineering strain rate at every simulation time step. In particular, to ensure accurate uniaxial loading condition, periodic simulation box along the structure width (perpendicular of the loading) was automatically adjusted using the NPT method to have negligible stress in this direction. In order to avoid any sudden bond stretching or void formation due to the loading conditions, the atomic positions at every step of the loading were rescaled according to the applied changes in the simulation box size. Virial stresses were calculated at each strain level, and were averaged over 1 ps intervals, to report the engineering stress-strain responses.

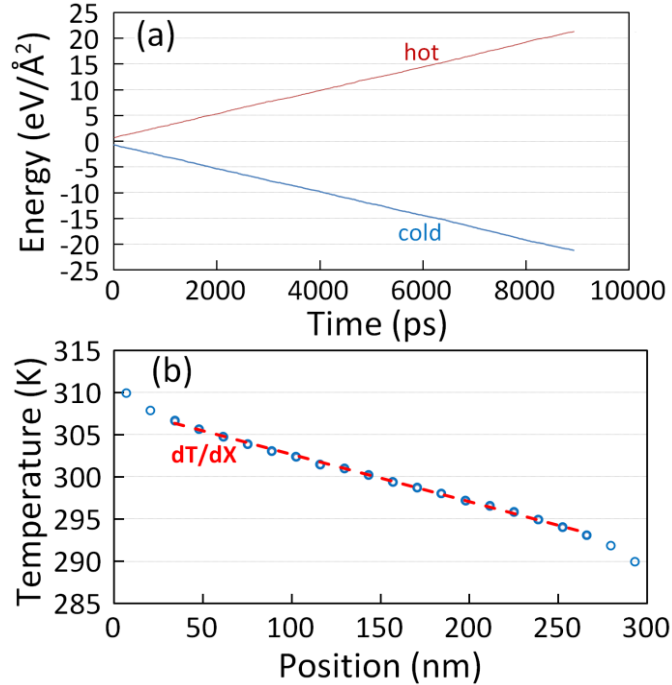

Fig. 2- (a) Applied energy curves to the atoms in the hot and cold reservoirs to keep the applied temperature difference constant during the simulations. (b) Averaged temperature profile along the sample present a linear relation (dT/dx), as shown by the dashed line.

We used non-equilibrium molecular dynamics (NEMD) method to predict the thermal conductivity of 2D $C_3N$ at the room temperature. The time increment of NEMD simulations was set to 0.5 fs and the periodic boundary conditions were also applied along the planar directions. In the NEMD approach, we first equilibrated the structures at the room temperature and zero stresses using the NPT method. Then atoms at the two ends were fixed and the system was further equilibrated using the Nosé-Hoover thermostat method (NVT). By neglecting the fixed atoms at the two ends, the rest of the simulation box was partitioned into 22 slabs and then a temperature difference of 20 K was applied only between the first and last slabs. In this step, the temperatures for these two slabs, so called hot (at 310 K) and cold (at 290 K) reservoirs were controlled at the adjusted values using the NVT method, while the rest of the structure (remaining 20 slabs) was simulated using the constant energy (NVE) method. In order to keep the applied temperature difference of 20 K along the sample, at every simulation time step an amount of energy was added to the atoms in hot reservoir and another amount of energy was removed from the atoms in the cold reservoir by the NVT method. After reaching to the steady-state heat transfer condition, a temperature gradient is established along the sample. The NEMD simulations were then performed for longer times, the calculated temperatures at every slabs were averaged and the energy values added or removed were recorded. Samples of the energy curves added and removed from the structure are plotted in Fig. 2a. Here, it is clear that the total energy of the system was accurately conserved and a constant heat-flux, $J_x$, was imposed on the structure. As it is shown in Fig. 2b, a linear temperature relation ($dT/dx$=constant) was also established along the sample. Based on the applied heat-flux, $J_x$, and the established temperature gradient, $dT/dx$, the thermal conductivity, $k$, of pristine $C_3N$ was calculated using the one-dimensional form of the Fourier law:

$$k = J_x \frac{dx}{dT} \qquad (1)$$

## 3. Results and discussions

Since the mechanical and thermal properties are directly related to the assumption for the thickness of the 2D $C_3N$, we first calculate the thickness using the DFT calculations. Here, van der Waals interactions were included using the semiempirical correction of Grimme [47], as it is implemented in VASP. In this case, we constructed bi-layer $C_3N$ structures with two different stacking sequences of AA and AB as they are shown in Fig. 3a and Fig. 3b, respectively. In the AA stacking, the in-plane position of atoms on the two layers are exactly the same. In the AB stacking, the in-plane atomic positions of the top layer are shifted such that some of the atoms are placed on the hollow center of the hexagonal lattices of the atoms on the bottom layer. After the geometry optimization and energy minimization, the distances between two $C_3N$ layers were obtained to be 3.3 Å and 3.2 Å, for the AA and AB stackings, respectively. We found that the total energy of the system with AB stacking is by around 0.1% lower than that with AA stacking which suggests that the AB stacking is more favorable. We therefore assume a thickness of 3.2 Å for the single-layer $C_3N$ to report the thermal conductivity. We note that the thickness of s-triazine-based graphitic carbon nitride (SgCN) was measured experimentally to be 3.28 Å [17], which is lower than the thickness of graphene. To provide a better vision concerning the stability of 2D $C_3N$, we next compare the energy of single-layer $C_3N$ with graphene and other 2D carbon-nitride structures that have been fabricated so far. For this purpose, we also included the energy of phagraphene [48], which has been reported to be the second 2D carbon allotrope with the lowest energy after the graphene. The results shown in Fig. 3c, reveal lower energy for the single-layer $C_3N$ and therefore may confirm its higher stability in comparison with other 2D carbon-nitride structures considered here.

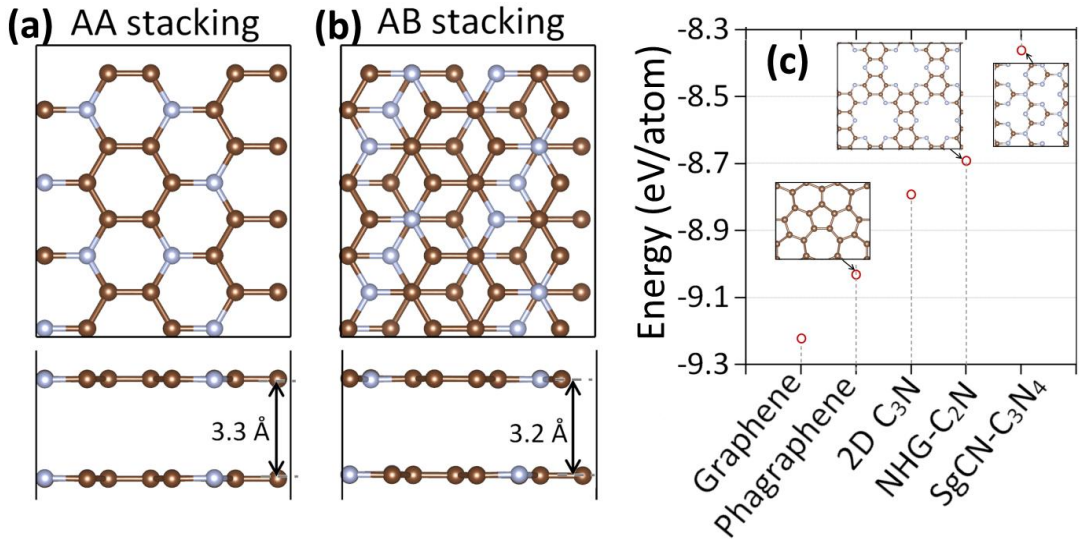

Fig. 3- (a) AA and (b) AB stacking of bi-layer 2D $C_3N$ after the energy minimizations using the DFT method. (c) Calculated total energy (eV/atom) of single-layer and pristine graphene, phagraphene [48], 2D $C_3N$, nitrogenated holey graphene (NHG) [26] and s-triazine-based graphitic carbon nitride (SgCN) [17] to compare their energy stability.

DFT results for the uniaxial stress-strain responses of defect-free and single-layer 2D $C_3N$, stretched along the armchair and zigzag directions are illustrated in Fig. 4. As it is shown, the stress-strain curve includes an initial linear relation, followed by a nonlinear response up to the ultimate tensile strength, where the maximum load bearing of the sheet is reached. By further loading after the ultimate tensile strength point, the stress suddenly drops which indicates that the failure has occurred in the specimen due to the bond rupture. Based on the results shown in Fig. 4, the linear parts of the stress-strain responses coincide closely for the both armchair and zigzag loadings. The slope of the initial linear part of the uniaxial stress-strain curve is equivalent with the elastic modulus. For single-layer $C_3N$, our DFT results accordingly predict elastic modulus of 341.4 GPa.nm and 339.7 GPa.nm, along the armchair and zigzag loading directions, respectively. Interestingly, our prediction for the $C_3N$ elastic modulus is only ~3% smaller than the elastic modulus of graphene, calculated to be 350.7 GPa.nm (1050 GPa) by Liu et al.[49]. For the initial linear part of the stress-strain curve, the ratio of the strain along the traverse direction ($\varepsilon_t$) with respect to the loading strain ($\varepsilon_l$) is acceptably constant and the Poisson's ratio can be obtained based on the $-\varepsilon_t/\varepsilon_l$ ratio. We found that the Poisson's ratio of single-layer $C_3N$ is around 0.155 and 0.14, along the armchair and zigzag directions, respectively. Since the elastic modulus of single-layer $C_3N$ are very close along the armchair and zigzag directions, one can conclude isotropic elastic response for $C_3N$ nanomembranes.

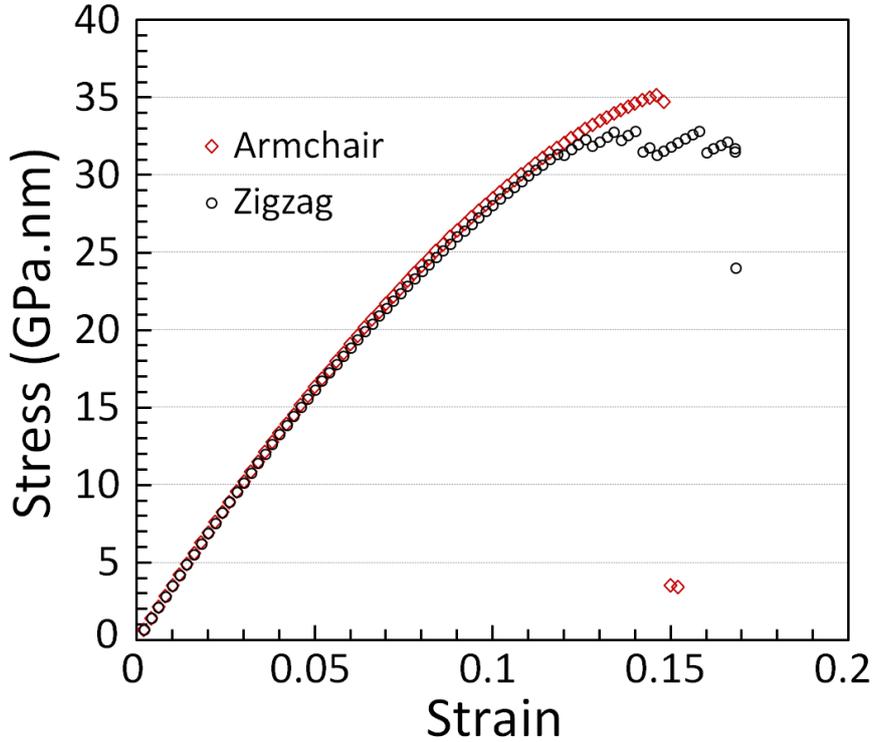

Fig. 4- First-principles density functional theory results for the uniaxial stress-strain responses of single-layer $C_3N$ along the armchair and zigzag directions.

Our DFT results shown in Fig. 4 reveal that the stress-strain responses of single-layer $C_3N$ elongated along the armchair and zigzag, match very closely up to the strain of 0.13. We found that the tensile strength of $C_3N$ sheets is however anisotropic and the ultimate tensile strength along the armchair is 35.2 GPa.nm, which is around 2.4 GPa.nm stronger than that stretched along the zigzag direction. Interestingly, our predictions for the tensile strength of single-layer $C_3N$ are only slightly below the tensile strength of pristine graphene, reported to be 36.7 GPa.nm (110 GPa) and 40.4 GPa.nm (121 GPa), by Liu et al.[49]. We should however note that as elaborately discussed in several studies [49-52], under different loading conditions and strain levels the dynamical instability may occur which may leads to materials failure at lower strains. The investigation of dynamical instability of 2D carbon-nitride allotropes including $C_3N$ under different loading conditions can be therefore an interesting topic for the future studies. The strain at the failure point is another important parameter for the mechanical response of a material, which indicates that how long a material can be stretched before a crack is formed in the lattice. Our DFT results show that along the zigzag direction, the $C_3N$ can withstand up to a strain level of 0.168, which is around 0.02 higher than that stretched along the armchair direction. For the pristine graphene, the failure strain along the armchair and zigzag were found to be 0.266 and 0.194, respectively, as reported in the work by Liu *et al.* [49]. This comparison reveals lower stretchability of 2D $C_3N$ in comparison with pristine graphene. Nevertheless, in comparison with graphene the mechanical response of 2D $C_3N$ is more isotropic.

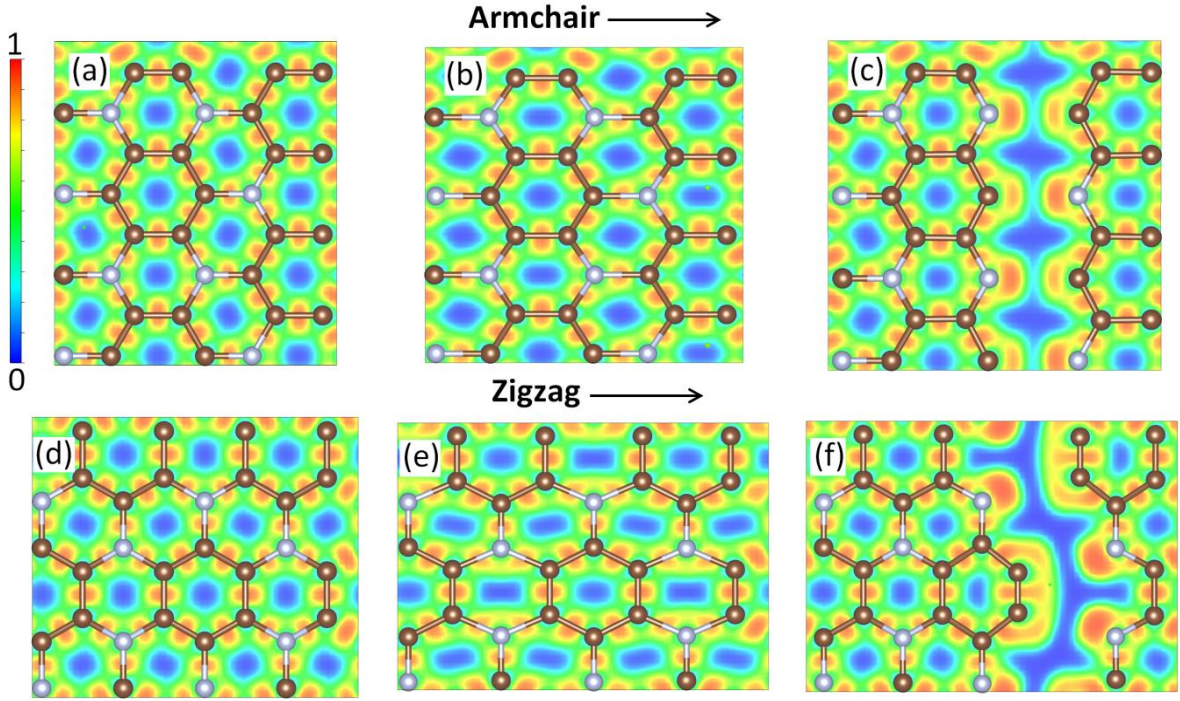

Fig. 5- DFT results for the deformation process of single-layer $C_3N$, stretched along the (a-c) armchair and (d-f) zigzag directions. The contours illustrate electron localization function (ELF), which has a value between 0 and 1, where 1 corresponds to perfect localization and ELF = 0.5 corresponds to the electron gas. We considered three different strain levels: (a and d) structure at the minimum energy, (b and e) at ultimate tensile strength point and (c and f) shortly after the rupture.

The DFT results for the deformation process of single-layer $C_3N$, stretched along the armchair and zigzag directions are compared in Fig. 5. We considered three different strain levels: structure at the minimum energy (under no loading), at ultimate tensile strength point and shortly after the rupture. In Fig. 5, we also plotted the electron localization function (ELF), in order to establish a connection between the electronic structure and the resulting mechanical responses. The high electron localizations occurring at the center of carbon-carbon and carbon-nitrogen bonds indicates the character of covalent bonds where the electrons are shared between two connecting atoms. The higher electron localization on a particular bond consequently illustrates higher rigidity of that bonds [53] and such that the comparison of electron localization can provide useful information for the evaluation of mechanical response and failure behaviour as well. Based on our DFT results for the relaxed structure and under no loading condition (Fig. 5a and Fig. 5b), homo-nuclear carbon-carbon bonds yield a higher localized electrons as compared with carbon-nitrogen. This observation can well explain the slightly lower elastic modulus of $C_3N$ in comparison with that of the graphene. This electron localization function for minimized structure also suggests that the initial bond breakage and the subsequent crack formation in the $C_3N$ nanomembranes occurs initially by the rupture of carbon-nitrogen bonds. This was confirmed by our DFT results for the ruptured samples (Fig. 5c and Fig. 5f), where it was observed that only the carbon-nitrogen bonds are broken and the homo-nuclear bonds are kept intact. In this way, for the both armchair and zigzag directions, the edges of the crack show zigzag pattern. For the uniaxial loading along the armchair, the deformation process is highly symmetrical (Fig. 5b). In this case, crack shows a straight line, exactly perpendicular of the loading. On the other side, for the loading along the zigzag direction, the deformation is not uniform (Fig. 5e) and the crack growth follows a more complicated path and this can explain the irregular stress pattern in the stress-strain curve, around the ultimate tensile strength point (Fig. 4). As it is shown in Fig. 5c and Fig. 5f, along the rupture edges the ELF that was initially maximum at the center of the bonds splits into two parts presenting a minimum between the two carbons and nitrogen atoms indicating the rupture of carbon-nitrogen covalent bonds.

Next, we shift our attention to the MD results. Using the MD simulations, one can take into account the dynamics factors such as the temperature effect and it is also possible to study the samples with larger number of atoms. To validate our MD modelling, we first compare the elastic modulus acquired by MD with that predicted by the DFT. Since in our DFT modelling the temperature effects are neglected, we performed the MD simulations of uniaxial tension at a low temperature of 2 K. Based on our MD results, the elastic modulus of single-layer $C_3N$ was predicted to be 333 GPa.nm and 336.7 GPa.nm along the armchair and zigzag directions, respectively. Interestingly, our MD predictions for the elastic modulus of $C_3N$ are only in 3% and 1% difference with our DFT results for the elastic modulus of 2D polyaniline along the armchair and zigzag, respectively. This considerably close agreement between the MD and DFT predictions, clearly confirms the accurate choice of potential functions in our MD modelling.

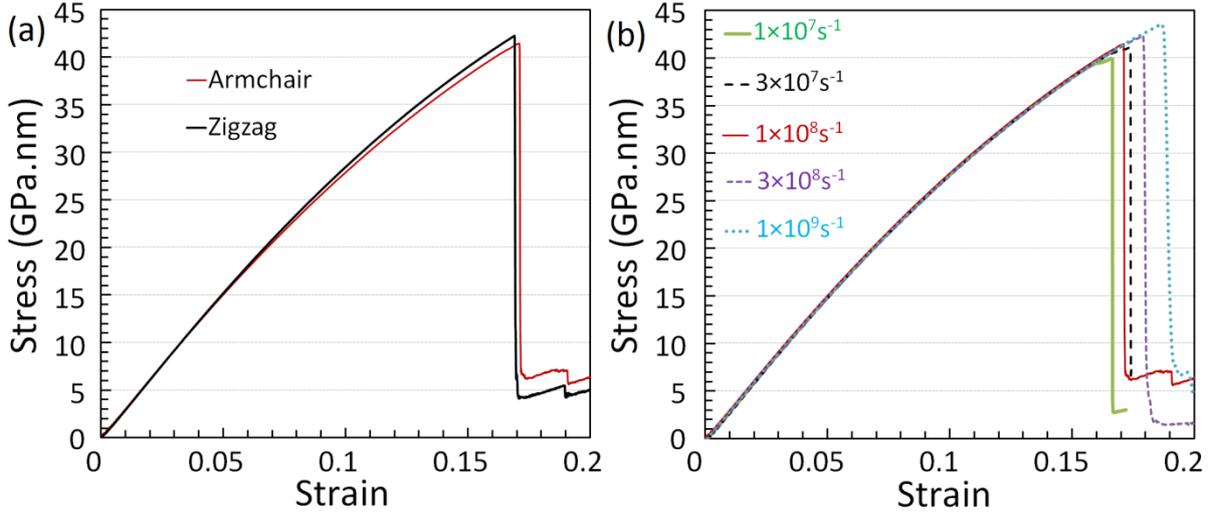

Fig. 6- Classical molecular dynamics results for the stress-strain response of single-layer $C_3N$ at the room temperature. (a) The stress-strain relations of samples elongated along armchair and zigzag with a strain rate of $1\times10^8$ s$^{-1}$. (b) Strain rate effect on the mechanical response of sample stretched along the armchair direction.

As it was discussed in detail in our recent study [54], we modified the cutoff of the optimized Tersoff [45] potential from 0.18 nm to 0.20 nm to simulate the mechanical response of pristine graphene. At the room temperature, our cutoff modified molecular dynamics model yields an elastic modulus of $960\pm10$ GPa and tensile strength of 132 GPa, for graphene, which match excellently with experimental results of $1000\pm100$ GPa and $130\pm10$ GPa [7] for the elastic modulus and tensile strength of pristine graphene, respectively. In our MD modelling in this study, we also considered the modified cutoff for the carbon-carbon interactions. Fig. 6a, illustrates the obtained stress-strain responses of 2D $C_3N$ at the room temperature. Here, we conducted the MD simulations employing a strain rate of $1\times10^8$ s$^{-1}$, since in our recent study [54] for the pristine graphene, with this loading condition we found excellent agreements for the predicted mechanical properties as compared with experimental results [7]. At the room temperature, our MD modelling predict an elastic modulus of around 313.2±1.5 GPa.nm and tensile strength of ~41-42 GPa.nm at corresponding failure strains of ~0.17. Our results show very close mechanical responses of single-layer $C_3N$ when stretched either along the zigzag or armchair directions, though it is slightly more stiff along the zigzag direction. In comparison with our MD results for the pristine graphene [54], the 2D $C_3N$ tensile strength is only around 7% smaller, which clearly reveals the ultra high mechanical response of single-layer $C_3N$. Interestingly, our predicted failure strain of ~0.17 by the MD simulations, is very close to our DFT result for the $C_3N$ sheet stretched along the zigzag direction. We should however remind that the strain rates applied in MD simulations are by several orders of magnitude faster than real experimental setups. To investigate the strain rate effect on the MD results for the 2D $C_3N$ mechanical properties, we performed the simulations at different strain rates and the acquired results are shown in Fig. 6b. As it can be seen, for different strain rates the stress-strain responses coincide up to the tensile strength point. As a general trend, by decreasing the strain rate the tensile strength decreases. We found that this trend is consistent for the loading along the both armchair and zigzag directions. The strain rate effect can also explain the overestimation of DFT result for the tensile strength by the classical MD simulations.

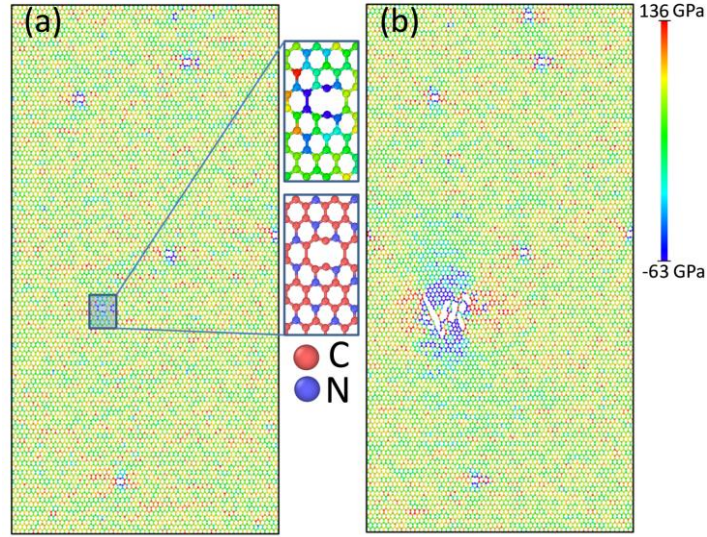

Fig. 7- MD results for the failure process of a single-layer $C_3N$ membrane, stretched with strain rate of $10^8 \, s^{-1}$ along the armchair direction at the room temperature, (a) at the ultimate tensile strength point, (b) shortly after the tensile strength point. The contours illustrate the atomic uniaxial stresses, plotted using the OVITO [79] package. In the stress contours presented here, all the stress values over the 136 GPa were shown with the similar colour. We assumed a thickness of 3.2 Å for the single-layer $C_3N$ in the stress calculations.

The deformation process of a single-layer $C_3N$ membrane, stretched along the armchair direction at 300 K is depicted in Fig. 7. For the $C_3N$ sheets stretched along the zigzag and armchair directions, we found that they extend uniformly and remain defect-free up to the strain levels close to the failure strain. For the $C_3N$ membrane stretched along the armchair direction, at strain levels close to the tensile strength point, few carbon-carbon bonds were broken (Fig. 7a inset) resulting in the formation of voids, consisting of 10 membered rings (Fig. 7a inset). The ultimate tensile strength reaches at a point in which the coalescence of the two previously formed voids occur (Fig. 7b), resulting in the formation of a crack that grows rapidly, leading to the sample rupture (Fig. 7b). Based on our classical MD simulations, single-layer $C_3N$ presents brittle failure mechanism at the room temperature. This conclusion is due to the fact that the defects formation and sample rupture occur at considerably close strain levels. In agreement with our DFT results, the edges of the formed crack along the sample show mainly the zigzag pattern. Nevertheless, the earlier ruptures of the carbon-carbon bonds are in contradiction with our first-principles predictions. Such an inaccuracy in our MD simulations can be corrected by modifying the carbon-nitrogen cutoff length, in the Tersoff potential parameters set. Our MD results show that because of the formation of the mono-atomic carbon chains during the rupture process, the crack growth in the single-layer $C_3N$ does not follow an straight line and such that the edges of the ruptured sample are not smooth.

Based on the experimental results, $C_3N$ nanomembranes present semiconducting electronic character, with a band gap of around 2.7 eV [27]. The DFT calculations however suggested that the single-layer $C_3N$ is a metallic conductor [27]. It is quite well-known that the DFT method based on the PBE functional underestimates the band-gap. HSE06 method provide more accurate results for the band-gap [55], we therefore also employed this hybrid functional to study the electronic DOS. We selected several single-layer $C_3N$ sheets at different strain levels with respect to the tensile strength point and then we calculated the total electronic DOS. Fig. 8, illustrates the acquired electronic DOS curves for the $C_3N$ sheet elongated along the armchair direction at different strain levels predicted using the (a) PBE and (b) HSE06 functionals. We found very similar trends for the single-layer $C_3N$ stretched along the zigzag direction. As it can be observed, for the sample under no loading condition, band-gaps of ~0.4 eV and ~1.1 eV are observable in PBE and HSE06 results, respectively. As expected, the PBE method underestimates the ban-gap predicted by the HSE06. According to our DFT results using the both functionals, the band-gap remains almost unchanged up to a high strain of $2/3\varepsilon_{uts}$. However, based on the PBE results for the structure at the ultimate tensile strength point, at the Fermi level the DOS is not zero which indicates metallic character. On the other hand for the same sample, HSE06 reveals a band-gap of ~1.0 eV which shows only an insignificant change in comparison with the samples under lower strain levels. We note that applying a high strain close the ultimate tensile strength point is not safe and may cause the material failure, which can be originated from a single defect in the lattice. Therefore our DFT results reveal limited possibility of the band-gap modification in single-layer $C_3N$ through applying the mechanical strains. Although our results match better with experimental results as compared with the earlier DFT study [27], we however emphasis that in the future studies more elaborated hybrid functionals such as GW [56–60] should be employed to discuss the electronic structure of $C_3N$ nanomembranes under different loading conditions.

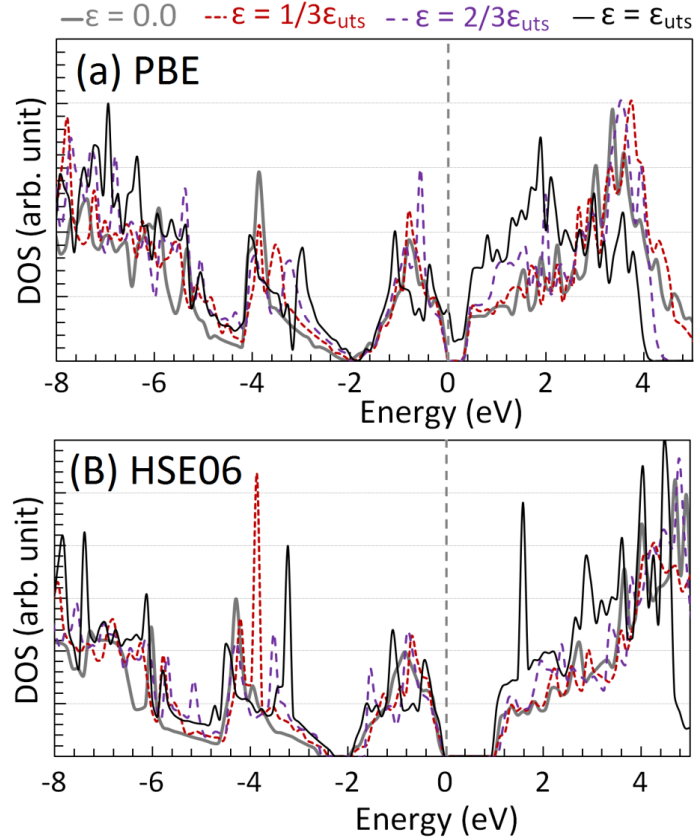

Fig. 8- Total electronic density of states (DOS) for single-layer $C_3N$ under uniaxial tensile loading along the armchair direction at different strain levels, ε, with respect to the strain at ultimate tensile strength, $\varepsilon_{uts}$.

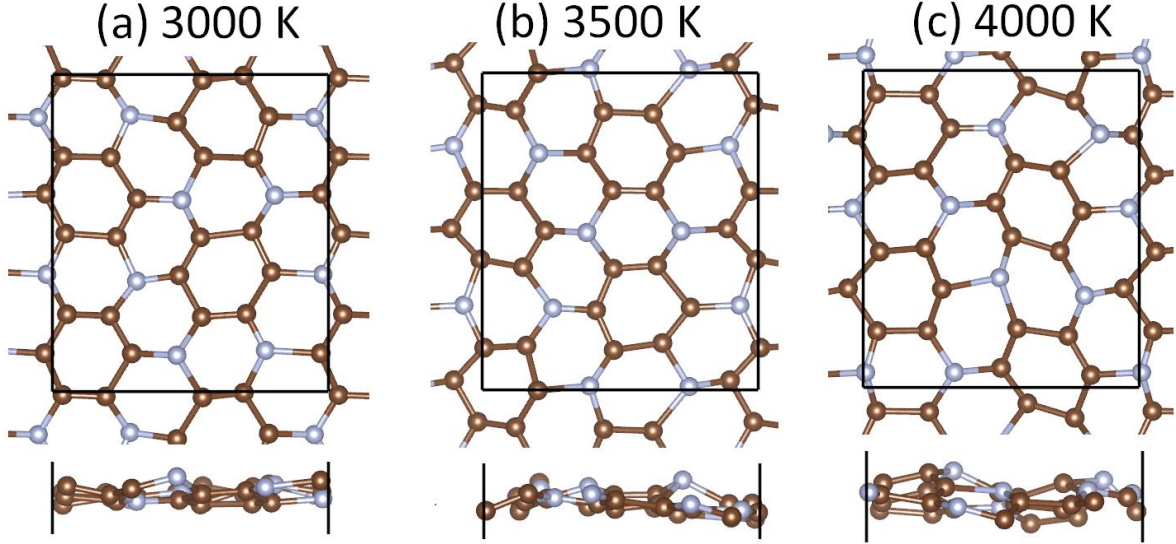

Fig. 9-Top and side views of final snapshot of the single-layer $C_3N$ structure after AIMD simulations for 10 ps.

Thermal stability of a material at high temperatures is among the appealing properties for the practical applications. We therefore studied the thermal stability of single-layer $C_3N$ at high temperature using the AIMD simulations for 10 ps. As it is shown in Fig. 9, the single-layer $C_3N$ remained intact at the end of the 10 ps simulation time at high temperatures of 3000 K to 4000 K. At higher temperatures than 4000 K, we found that it was partly disintegrated and at 5000 K it was completely disintegrated. Our AIMD results suggest that single-layer $C_3N$ owing to its ultra high stiffness, can withstand at high temperatures like 4000 K, which is only ~12% lower than the melting point of graphene, reported to be around 4510 K [61]. We note that as simulated in our previous AIMD study [29], the nitrogenated holey graphene cannot withstand at temperatures like 4000 K which further confirms the higher thermal stability of 2D $C_3N$.

For the evaluation of the single-layer $C_3N$ thermal conductivity, we performed the simulations for samples with different lengths to explore the length dependency. In Fig. 10, we plot the NEMD results for the length effect on the predicted thermal conductivity of $C_3N$ along the armchair and zigzag directions at room temperature. For the small lengths, the thermal conductivity sharply increases by increasing the sample length, which is related to the ballistic thermal transport. However, the increase in the thermal conductivity values is considerably slowed down for the samples with sizes larger than 200 nm, implying that the thermal transfer approaches the diffusive heat transfer. The thermal conductivity of single-layer $C_3N$ with infinite length, $k_\infty$, can be calculated by the extrapolation of the NEMD results for the samples with finite lengths, $k_L$ [62]. The length dependence, $L$, of the thermal conductivity can be described by [62,63]:

$$\frac{1}{k_L} = \frac{1}{k_\infty}\left(1 + \frac{\Lambda}{L}\right) \quad (2)$$

Here, the $\Lambda$ is the effective phonon mean free path (EMFP). Based on the Eq. 2, one can find that for the $L=\Lambda$, the $k_L = k_\infty/2$, in this way, the EMFP corresponds to a length at which the system yields a thermal conductivity, half of the length independent thermal conductivity, $k_\infty$, [29,64–66]. By fitting curves to the NEMD data points, the diffusive or in another word length independent phononic thermal conductivity of single-layer and free-standing $C_3N$ along the armchair and zigzag directions at room temperature are calculated to be 810±20 W/mK and 826±20 W/mK, respectively. The EMFP along the armchair and zigzag directions were also predicted to be around 75.5 nm and 77 nm, respectively. Likely to our predictions for the elastic modulus of single-layer $C_3N$, its thermal conductivity along the armchair and zigzag directions are also considerably close. This observation suggests convincingly isotropic elastic and thermal conductivity response along the $C_3N$ nanomembranes.

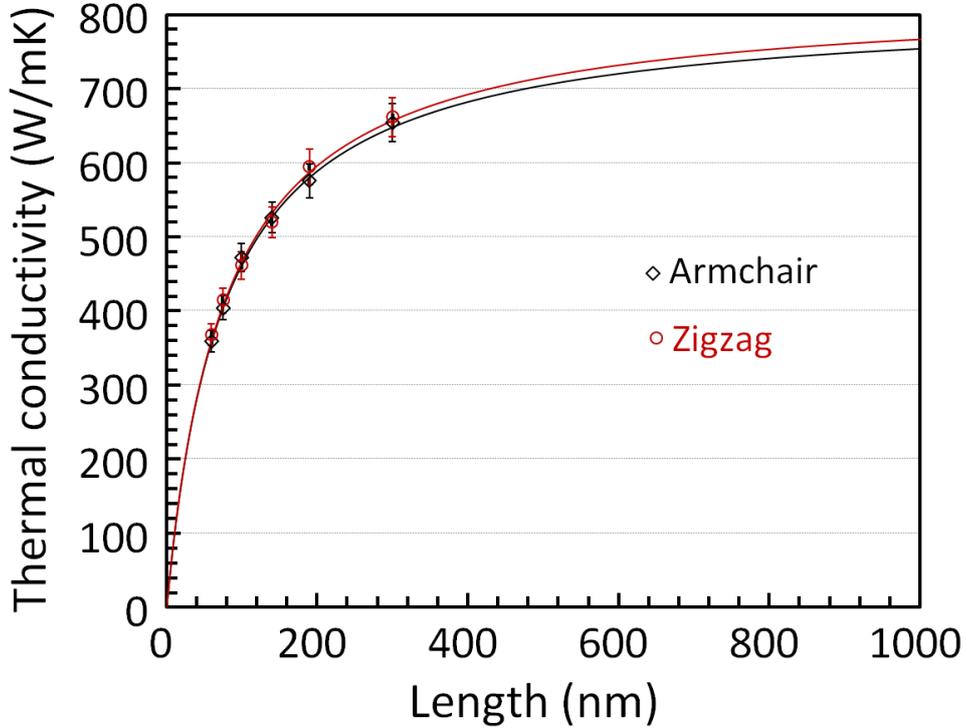

Fig. 10-The NEMD results for the length effect on the thermal conductivity of single-layer $C_3N$ along the armchair and zigzag directions calculated at 300 K. The solid lines illustrate the best fits to the NEMD data points. We assumed a thickness of 3.2 Å for the single-layer $C_3N$.

It is worthy to note that the thermal conductivity of triazine-based graphitic carbon structures and nitrogenated holey graphene were predicted to be 3.5-7.6 W/mK [67] and 64.8 W/mK [29], respectively. Interestingly, in comparison with these 2D carbon nitride structures synthesized so far, the thermal conductivity of 2D $C_3N$ is more than at least an order of magnitude stronger. By neglecting the graphene, the thermal conductivity of 2D $C_3N$ is also clearly higher than that measured along other 2D materials such as black phosphorene, 26.1-64.6 W/mK [68], molybdenum disulfide 34.5 ± 4 W/mK [69] and hexagonal boron-nitride, 250-600 W/mK [70–72]. It should be however taken into the consideration that using the classical NEMD simulations one cannot account for electronic thermal transport. As elaborately discussed in the recent first-principles study by Kim *et al.* [73], for the doped graphene the electronic contribution to the thermal conductivity can be around 10%. Since 2D $C_3N$ can be considered as a nitrogen doped graphene, owing to electronic contribution one can accordingly expect higher thermal conductivity for the real samples as those we predicted using the classical NEMD simulations. We should remind that because of the high electrical conductivity of graphene, the polymer nanocomposites made from graphene fillers are also probably electrical conductors which may cause undesirable effects in electronic devices. Semiconducting electronic property along with the high thermal conductivity of $C_3N$ nanomembranes are desirable characteristics and such that their polymer nanocomposites can present high thermal conductivity [74–78] and at the same time remain electrically insulators. In response to the thermal management concerns in various applications, particularly in nanoelectronics, $C_3N$ nanofilms can be considered as promising candidates. Ultra high mechanical properties of $C_3N$ nanosheets, also mean that the polymer nanocomposites made from $C_3N$ nanofillers may also present considerably enhanced mechanical responses.

## 4. Conclusions

Most recently, 2D polyaniline crystals with $C_3N$ stoichiometry and semiconducting properties have been successfully synthesized via the direct pyrolysis of hexaaminobenzene trihydrochloride single crystals in solid state. 2D polyaniline lattice is analogous to that of the pristine graphene, with an ordered pattern of nitrogen atoms substituting the native carbon atoms. We conducted extensive atomistic modelling to explore the mechanical properties and thermal conductivity of single-layer and free-standing $C_3N$. For single-layer $C_3N$, our first-principles density functional theory (DFT) results predict remarkably high elastic modulus of ~341

GPa.nm, tensile strength of ~33-35 GPa.nm and failure strains of 0.148-0.168, depending on the loading direction. Classical molecular dynamics (MD) simulations performed at a low temperature of 2 K, predicts the elastic modulus of free-standing $C_3N$ in a maximum 3% difference with the DFT results, which verifies the accuracy of our classical modelling. MD simulations for the uniaxial tensile simulation of single-layer $C_3N$ at the room temperature, predict an elastic modulus of around 313.2±1.5 GPa.nm. Electron localization function analysis reveals a higher localized electrons for homo-nuclear carbon-carbon bonds as compared with carbon-nitrogen bonds. Based on the DFT results, it was shown that the initial bond breakage and the subsequent crack formation in the $C_3N$ nanomembranes occurs only by the rupture of carbon-nitrogen bonds. DFT calculations suggest the limited possibility of the band-gap modification in the single-layer $C_3N$ through applying the mechanical strains. Ab initio molecular dynamics simulations highlights that single-layer $C_3N$ owing to its ultra high stiffness, can withstand at high temperatures like 4000K. The length independent phononic thermal conductivity of single-layer and free-standing $C_3N$ at room temperature was calculated to be 810-826 W/mK, using the classical non-equilibrium molecular dynamics method. We nevertheless expect higher thermal conductivity for 2D $C_3N$, because of the fact that in our classical modeling the electronic contribution to the thermal conductivity was not taken into the consideration. Our modelling results suggest convincingly isotropic elastic and thermal conductivity response along the $C_3N$ nanomembranes. After the graphene with the highest known thermal conductivity, the thermal conductivity along the free-standing $C_3N$ was found to be considerably stronger than other 2D materials. Our investigation confirms the outstanding thermal stability and ultra high mechanical strength and thermal transport along the single-layer and free-standing $C_3N$. By taking into consideration that $C_3N$ nanofilms are semiconducting, they have promising potentials to compete not only with graphene but also with other 2D materials for various applications, particularly in nanotransistors, fabrication of polymer nanocomposites with superior thermal and mechanical response and thermal management of electronics and energy storage devices. We therefore hope that the insight provided by our study can be useful for the future practical applications of $C_3N$ nanomembranes.

Acknowledgment

The first author greatly acknowledges the financial support by European Research Council for COMBAT project (Grant number 615132).